\documentclass[conference]{IEEEtran}
\IEEEoverridecommandlockouts
\usepackage{cite}
\usepackage{amsmath,amssymb,amsfonts}
\usepackage{algorithmic}
\usepackage{graphicx}
\usepackage{textcomp}
\usepackage{xcolor}
\usepackage{flushend}

\def\BibTeX{{\rm B\kern-.05em{\sc i\kern-.025em b}\kern-.08em
    T\kern-.1667em\lower.7ex\hbox{E}\kern-.125emX}}

\newcommand{\zb}{\mathbf{z}}

\newcommand{\Zb}{\mathbf{Z}}

\begin{document}

\title{Intermittent Information-Driven Search
 for Underwater Targets}

\author{\IEEEauthorblockN{Branko Ristic}
\IEEEauthorblockA{\textit{School of Engineering} \\
\textit{RMIT University}\\
Melbourne, Australia \\
branko.ristic@rmit.edu.au}
\and
\IEEEauthorblockN{Alex Skvortsov}
\IEEEauthorblockA{\textit{Maritime Division} \\
\textit{Defence Science and Technology}\\
Melbourne, Australia \\
alex.skvortsov@dst.defence.gov.au}
}

\maketitle

\begin{abstract}
The problem is area-restricted search for targets using an autonomous mobile sensing platform. Detection is imperfect: the probability of detection depends on the range to the target, while the probability of false detections is non-zero. The paper develops an intermittent information-driven search strategy, which combines fast and non-receptive displacement phase (ballistic phase) with a slow displacement sensing phase. Decisions where to move next, both in the ballistic phase and the slow displacement phase, are information-driven: they maximise the expected information gain. The paper demonstrates the efficiency of the proposed strategy in the context of a search for underwater targets: the searcher is an autonomous amphibious drone which can both fly and land or takeoff from the sea surface.
\end{abstract}


\section{Introduction}

Searching strategies for finding targets using appropriate sensing
modalities are of great importance in many aspects of life. In the context of national security, there could be a need to find a source of hazardous emissions \cite{ristic_search_2010,ristic2016study,hutchinson2018information}. Similarly, rescue and recovery missions may be tasked with localising a lost piece of equipment that is emitting weak signals \cite{haley_stone_80}. Biological applications include, for example, protein searching for its specific target site on DNA \cite{HalfordSE2004HdsD}, or foraging behaviour of animals in their search for food or a mate \cite{albatros_96,fauchald2003using}. The objective of search research \cite{search_reserach} is to develop optimal strategies for localising a target in the shortest time (on average), for a given search volume.

The earliest theoretical studies of search strategies were conducted during WWII for the US navy. The goal was to design the most efficient flight paths for an aircraft in its search for enemy submarines~\cite{koopman1946search,champagne_03}. These classical approaches were concerned with systematic search, resulting in predetermined (deterministic) paths, such as the parallel sweep or the Archimedean spiral~\cite{haley_stone_80,bernardini2017combining}.
The search patterns of animals, on the contrary, are random rather than deterministic. An explanation for this phenomenon is that an event, such as a detection (false or true), changes the strategy and hence the behaviour of the searcher. Subsequent changes of strategy manifest themselves as a random-like motion pattern. Most of the current research into search strategies is towards the mathematical modelling and explanation of random search patterns~\cite{search_reserach,Benichou_2011,hutchinson2018information}.

By studying the GPS data of albatrosses,
it was discovered that search patterns of these birds consist of the segments whose lengths are random
draws from the Pareto-L\'{e}vy distribution \cite{albatros_96}. This discovery led to
several papers demonstrating that the so-called L\'{e}vy walks/flight are the optimal search strategy for foraging animals (deer, bees, etc), resulting in fractal geometries of search paths.

Search motion patterns, however, seem to depend on the ratio between
the search domain and the sensing range (or the density of targets).
Humpries {\em et al.} \cite{humpries_10} demonstrated that L\'{e}vy
behavior occurs only in environments where the targets (prey) are sparsely
distributed, while Brownian motion is optimal if the targets (prey) is
abundant.
An alternative to L\'{e}vy strategies is the \emph{intermittent search}: a
combination of a fast and non-receptive displacement phase (long jumps
within the search domain, with no sensing) with a slow search phase
characterised by sensing and reaction \cite{kramer_01}. B\'{e}nichou
{\em et al.} provide both a theoretical study and experimental data
verification of  intermittent search \cite{benichou_06, Benichou_2011}. In their
terminology, the fast relocation phase is referred to as the {\em
ballistic} flight with constant velocity $v$ and random direction.
The slow sensing/detection phase is modelled as either a motionless
wait or a {\em diffusive displacement} with diffusion coefficient
$D$. The optimal average duration of the two displacement phases is
derived as a function of the radius of the search circle, the radius
of the sensing circle, and the ratio of velocities $D/v$.

B\'{e}nichou {\em et al.} studied intermittent search without taking into account the information gathered by sensing during the search. Vergassola {\em at al.}
\cite{vergassola_07} proposed such a search strategy (referred as {\em infotaxis}). This  strategy selects the motion option that will maximise the expected rate of the
information gain. Information driven search by infotaxis made a huge impact on the research community; for a recent review see \cite{hutchinson2018information}. Vergassola {\em at al.} considered information driven search only in the slow sensing/detection phase.

This paper proposes an information-driven intermittent search strategy. In this strategy, displacement decisions (i.e. where to move next), both in the ballistic phase and the diffusive displacement phase,  are based on maximisation of the expected information gain. The proposed search strategy presented in the context of an autonomous amphibious drone searching for underwater targets.

The paper is organised as follows. Sec. \ref{s:prob} presents a mathematical formulation of the problem. Sec. \ref{s:strategy} describes the proposed intermittent information-driven search strategy. Numerical evaluation and comparisons are given in Sec. \ref{s:num}. Finally, the conclusions from this study are drawn in Sec. \ref{s:conc}.

\section{Problem formulation}
\label{s:prob}

The context is autonomous search for underwater targets.  The searching platform is an autonomous unmanned vehicle (UAV), which can both fly and float/takeoff from the sea surface. While the vehicle is floating on the water surface, it is in a sensing mode. Sensing is carried out for the purpose of collecting detections/measurements of nearby underwater targets (if present), using for example a hydrophone array. The detection process is naturally imperfect: the probability of detection is monotonically decreasing with the range to the target; in addition, there is a non-zero probability of false detections. The assumption is that the  UAV can travel short distance while it is on the water surface. This type of motion may be occasionally required in order to further investigate (and reduce uncertainty about) some detections. The flying mode of the vehicle corresponds to the ballistic phase of the intermittent search. It results in a fast displacement, but during this phase, sensing is not performed.

The search is carried out in a two-dimensional  area $\mathcal{S}$, discretised into a square lattice (grid) of size $b$, consisting of $C=b^2\gg 1$ cells. The grid period (the distance between the neighbouring cells) is $R_0\ll b$. The coordinates of $m$th cell of the grid are assumed known, and denoted $\mathbf{g}_m=(x_m,y_m)$. The set of all grid-cell coordinates is denoted $\mathbf{G} = \{\mathbf{g}_m; m=1,\dots,C\}$.

 Suppose a target is located in one of the cells of the grid. Following~\cite{krout_09}, the presence or absence of a target in the $m$th cell of the grid ($m \in \{1,\dots,C\}$ at time $k$ is modelled by a Bernoulli random variable $\delta_{k,m}\in\{0,1\}$, where by convention $1$ denotes that a target is present.

When the searcher is in the sensing mode, it collects at time $k$ a set of detections $\Zb_k=\{\zb_{k,1},\dots,\zb_{k,m_k}\}$ from the environment (within its sensing volume). Each detection consists of a range and azimuth measurement from the searcher location to the perceived target. The searcher can move only on grid-cell locations, that is, its location at time $k$ is $\mathbf{p}_k\in \mathbf{G}$. Moreover, $\mathbf{p}_k$ is assumed known; it can be measured for example with an onboard GPS receiver.

The probability of detection of a true target located in cell $m\in\{1,\dots,C\}$, considering that the searcher is in position $\mathbf{p}_k = (x_k,y_k)$, is denoted $P_d^{k,m}$.  Similarly, the probability of false alarm is denoted $P_{fa}^{k,m}$. Both $P_d^{k,m}$ and $P_{fa}^{k,m}$  can be modelled in an arbitrary manner. Let us adopt the probability of detection to be a function of the range (distance) $r^{k,n}$ between the $n$th grid-cell, whose coordinates are specified by $\mathbf{g}_n \in \mathbf{G}$ and the searcher position $\mathbf{p}_k$, i.e. $r^{k,m} = \| \mathbf{p}_k - \mathbf{g}_m \|$. In this way, if a target is in a cell at a short distance from the searcher, its detection probability will be high, and vice versa. The mathematical model of the probability of detection is adopted as
\begin{equation}
P_d^{k,m} = \exp \left( - r^{k,m} / a \right).
\label{e:prob_det}
\end{equation}
The sensing area is characterised by the constant $a$, which typically depends on the sensor and environment. We assume $a$ is known and specified in terms of grid period $R_0$.
With this specification, the probability of detecting a target at a distance of $r=3a$ is approximately $0.05$. Assuming $360^\circ$ coverage, the sensing area $\mathcal{L}_k$ can be seen as a circular area of radius $3a$. The spatial distribution of false alarms is assumed to be uniform over $\mathcal{L}_k$ and homogeneous over $\mathcal{S}$. The number of false detections in $\mathcal{L}_k$ is modelled by the Poisson distributed with the mean rate $\lambda$. The measured range and azimuth to the target are assumed to be affected by additive zero-mean Gaussian noise.

When the searcher is in the ballistic phase of the intermittent search, it moves with speed $V_0$. When it is the diffusion phase, it is either static, or moves from one cell to a neighbouring cell (on the sea surface) with a small speed $v_0 < V_0$. The searcher requires $\tau_0$ seconds to acquire a measurement in diffusion (sensing) phase.

Given the search grid $\mathbf{G}$, plus the specification of the sensor and environment ($P_d^{k,m}$, $P_{fa}^{k,m}$, parameters $a$, $V_0$, $v_0$), the problem is to design a search strategy which would minimise the average search time.

\section{Search Strategy}
\label{s:strategy}

\subsection{Threat map and information gain}

Let us first introduce the posterior probability of target presence in the $m$th cell at time $k$ after processing the sequence $\Zb_{1:k}:=\Zb_1,\cdots,\Zb_k$ of detection sets.
 This probability is defined as $P_{k,m} = Pr\{\delta_{k,m}=1|\Zb_{1:k} \}$, where $\delta_{k,m}$ is a Bernoulli random variable, introduced in Sec.~\ref{s:prob}. The probability map, also known as the threat-map, is then a collection $\mathbb{P}_k = \{P_{k,m}; m=1,\dots,C\}$. The threat map is updated using the Bayes' rule as follows. Given the threat map $\mathbb{P}_{k-1}$ and a detection set at time $k$, that is $\Zb_k$, the probability $P_{k,m}$ is computed as~\cite{krout_09}
\begin{equation}
  P_{k,m} = \frac{(1-P_d^{k,m})P_{k-1,m}}{(1-P_d^{k,m})P_{k-1,m} + (1-P_{fa}^{k,m})(1-P_{k-1,m})}
  \label{e:tm1}
\end{equation}
if none of the detections in $\Zb_k$ falls into the $m$th cell. If, on the contrary, a detection is received in the $m$th cell, then the update equation is
\begin{equation}
  P_{k,m} = \frac{P_d^{k,m}P_{k-1,m}}{P_d^{k,m}P_{k-1,m} + P_{fa}^{k,m}(1-P_{k-1,m})}.
  \label{e:tm2}
\end{equation}
Initially, that is before any sensing at $k=0$, the threat map is set to $P_{0,m}=\frac{1}{2}$, for $m=1,\dots,C$. In this work we only consider static targets, but for moving targets a diffusion process can be applied to every cell in the threat map $\mathbb{P}_{k-1}$ just before the update time $k$~\cite{krout_09}.

\begin{figure}[htb]
  \centerline{\includegraphics[width=4.4cm]{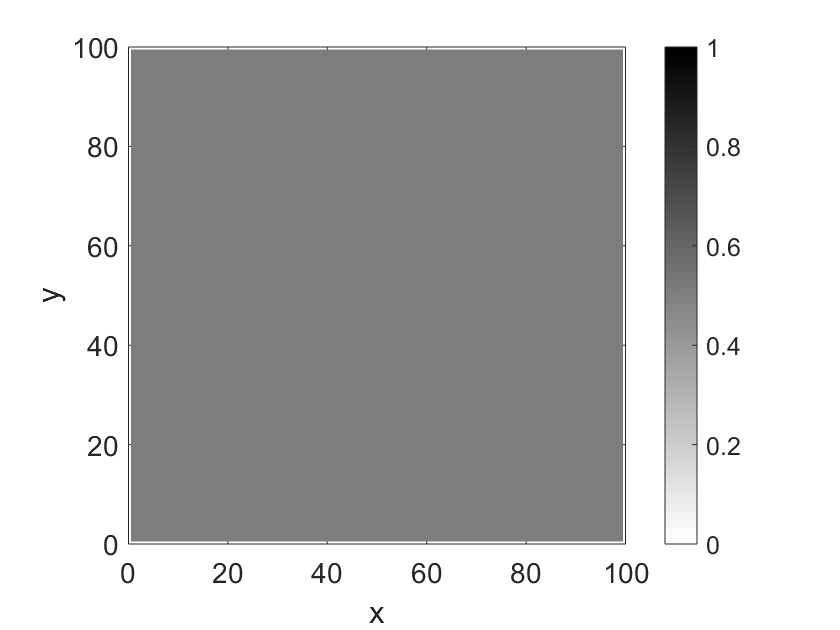} \includegraphics[width=4.4cm]{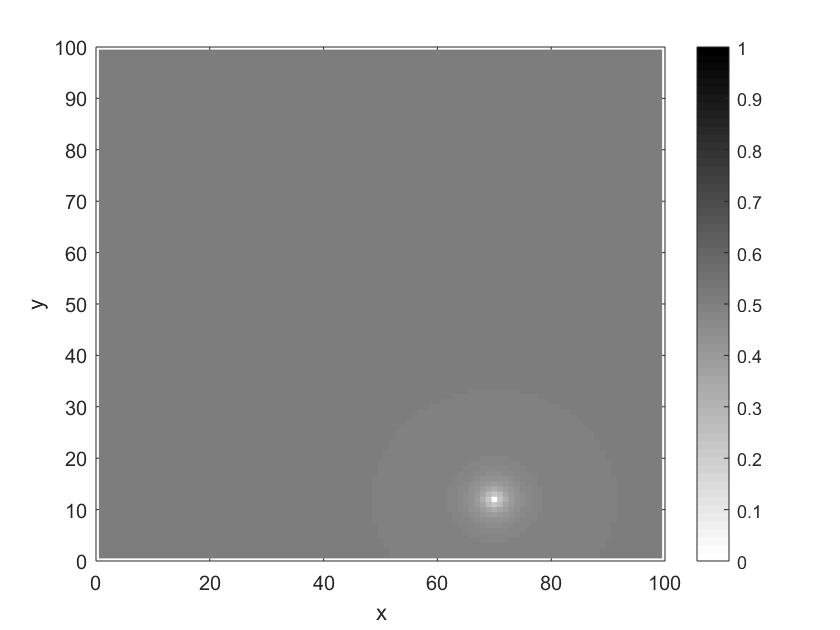} }
  \centerline{\footnotesize (a)\hspace{4cm}(b)\hspace{0.5cm}}
  \centerline{\includegraphics[width=4.4cm]{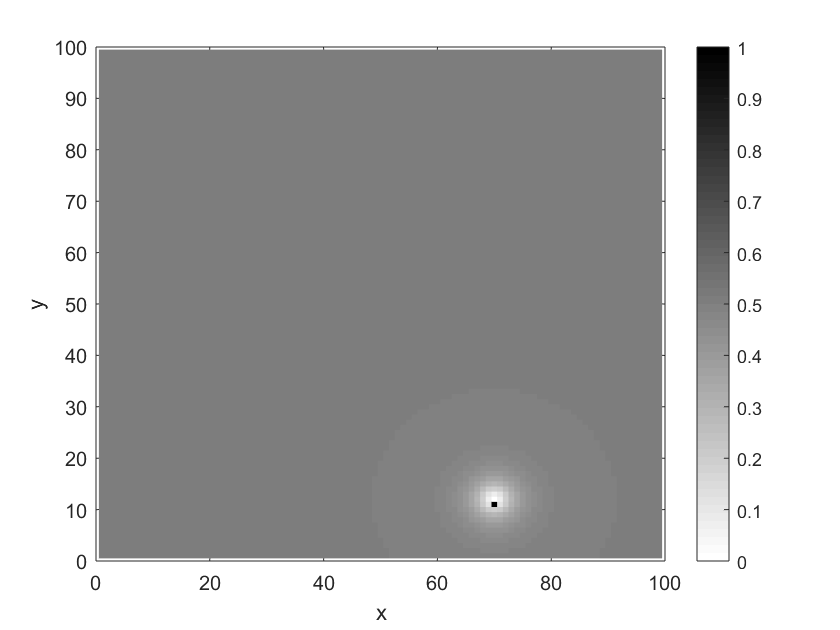} \includegraphics[width=4.4cm]{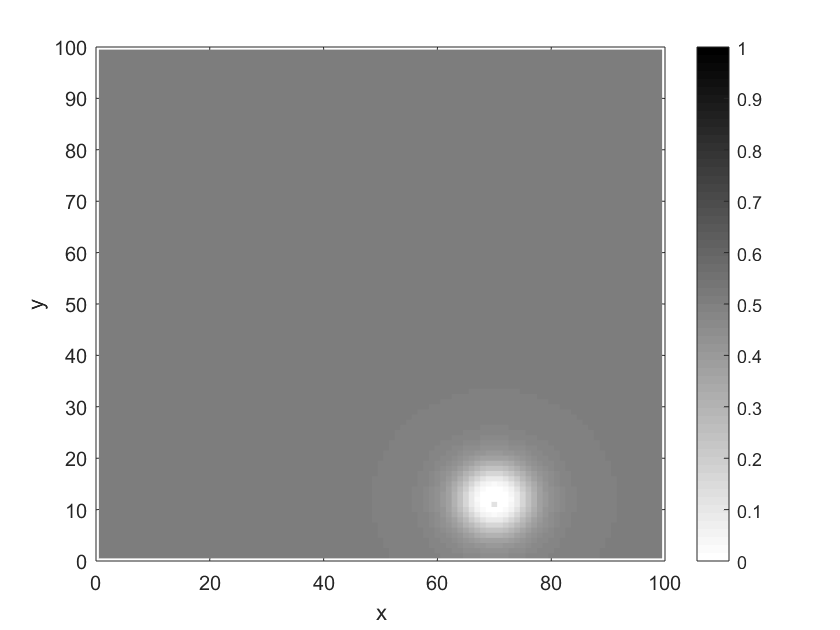} }
  \centerline{\footnotesize (c)\hspace{4cm}(d)\hspace{0.5cm}}
  \caption{Threat map evolution with a static searcher at coordinates $(70,12)$: (a) $k=0$; (b) $k=1$, and $\Zb_1=\emptyset$; (c) $k=1$ and $\Zb_2$ contains a single false detection in the cell at coordinates $(70,11)$; (d) $k=8$, with $\Zb_3=\cdots\Zb_8=\emptyset$}
  \label{f:1}
\end{figure}

An illustration of the threat map and its Bayes update using measurement sets $\Zb_{1:k}$ is given in Fig.~\ref{f:1}. The search grid consists of $C=100\times 100$ cells, with a period $R_0=1$ arbitrary units (a.u.). The searching platform is placed at the cell with coordinates $(70,12)$ a.u. and its sensor parameter is $a=3R_0$. The target is in the cell with coordinates $(35,60)$ a.u., that is at the distance where the searcher cannot detect it.  Fig.~\ref{f:1} displays the threat map at $k=0,1,2$ and $k=8$, using the gray-scale intensity plots. At $k=1$, the measurement set is empty, i.e. $\mathbf{Z}_1=\emptyset$. The threat map shown in Fig.~\ref{f:1}.(b) is obtained by updating the initial threat map using (\ref{e:tm1}) for all $m=1,\dots,C$. The white area at and near the position of the searcher, indicates a low (almost zero) probability that the target is present in those cells. Fig.~\ref{f:1}.(c) shows the threat map after receiving $\Zb_2$ which contains a false detection at the cell with coordinates $(70,11)$. This cell of the threat map was updated using (\ref{e:tm2}). Finally, Fig.~\ref{f:1}.(d) displays the threat map at $k=8$, with $\Zb_3=\cdots=\Zb_8=\emptyset$. The threat map is now characterised by a larger area with zero probability of target presence around the searcher position. Moreover, even the probability in the cell that received a false detection at $k=2$ is also very low.

By staying longer in the same position, the white area around the searcher position would grow only up to a certain saturation level, determined by  the probability of detection as a function of distance. The measurements received after reaching this saturation level would be increasingly uninformative.

\subsection{The reward function}

The search can be seen as a repetitive cycle of {\em sensing}, {\em threat map update} and {\em decision making} over the action space, where each action determines where to collect the next measurement. The reward function quantifies the benefit of an action. Thus, autonomous search is a form of reinforcement learning, where the searcher chooses the action that will maximise the reward.

The reward function is adopted as the reduction in entropy of the threat map, and defined as
\begin{equation}
  \mathcal{R}_k(\alpha)  = H_{k-1} - \mathbb{E}\{H_k(\alpha)\} \label{e:Rew}
\end{equation}
where
\begin{itemize}
    \item $\alpha\in\mathcal{A}_k$ is an action from the action set $\mathcal{A}_k$,
    \item $H_k$ is the entropy of the threat map $\mathbb{P}_k$, defined as
\begin{align}
  H_k = -  \frac{1}{C} \sum_{m=1}^{C} & \big[ P_{m,k}\,\log_2 P_{m,k} + \nonumber \\
  & (1-P_{m,k})\,\log_2 (1-P_{m,k})\big]
  \label{e:H}
\end{align}
Here $\mathbb{E}$ is the expectation operator with respect to $p(\Zb_k(\alpha)|\Zb_{1:k-1})$.
\end{itemize}
The expectation operator is necessary because the decision has to be made prior to collecting $\Zb_k$ (the detection set at time $k$).

Note that by setting the initial threat map to be $P_{0,m}=1/2$ for $n=1,\dots,C$, the initial entropy according to (\ref{e:H}) is $H_0=1$.

\subsection{Intermittent search}

The search objective, to find and localise the target in the shortest time, it is driven by two conflicting demands:  {\em exploration} and {\em exploitation}. The exploration demand is forcing the searcher to constantly move and thus investigate as much of the search volume as possible. Since the detection probability and the measurement accuracy are inversely proportional to the distance, the exploitation demand is urging the searcher to stay longer in one place. This helps to determine with certainty if a detection is false or true and improves the localisation accuracy.  The balance between exploration and exploitation exposes the universal dilemma in decision making: should I stay or should I go?~\cite{cohen2007should}.

Intermittent search strategy \cite{benichou_06,Benichou_2011}  was proposed as a balance between exploration and exploitation. Exploitation is carried out while the searcher is in the diffusion phase (with no, or very limited, motion). Exploration corresponds to the ballistic flight phase. The questions are then: What should be the duration of each of the two alternating phases of the intermittent search? Where the searcher should fly to in the ballistic phase?

We propose that the duration of any phase in intermittent search is a random draw from the exponential distribution \cite{Benichou_2011}, i.e.
\begin{equation}
t \sim \frac{1}{\tau}\exp\left(-\frac{t}{\tau}\right)
\label{e:exp}
\end{equation}
where the parameter $\tau$ is
\begin{equation}
    \tau = \begin{cases}
         \tau_d, & \text{if diffusion phase}\\
         \tau_b, & \text{if ballistic phase.}
    \end{cases}
\end{equation}

The ballistic time parameter $\tau_b$ is defined by
\begin{equation}
    \tau_b=\gamma\frac{a}{V_0}
\end{equation}
where $V_0$ and $a$ were introduced in Sec. \ref{s:prob} and $\gamma$ is a is a numerical factor dependent on the search area geometry \cite{Benichou_2011}:
\begin{equation}
    \gamma = [\ln(b/a)-1/2]^{1/2}.
\end{equation}
Note that the value of $\gamma$ slowly increases with the ratio $b/a$ (as a $\ln(\cdot)$). Furthermore, if $b\gg a$ then $\gamma > 1$; if $b=1.65\,a$, then $\gamma=0$; if $b=4.5\,a$, then $\gamma=1$. Since the speed of the ballistic flight is constant,  the length of the ballistic flight $L_0=t\,V_0$ is random because of (\ref{e:exp}).

The diffusion time parameter $\tau_d$ is determined as follows. Note that after collecting one measurement, the probability that the searcher detects a target within the range $r\leq L_0$ is from (\ref{e:prob_det}):
\begin{equation}
    Pr(r=L_0) = \exp(-L_0/a) \ll 1.
\end{equation}
After collecting $n$ measurements, the probability that the searcher does not detect the target is
\begin{equation}
    P_n = [1-\exp(-L_0/a)]^n.
\end{equation}
Let us assume that the searcher should jump out of search area with radius $r\leq L_0$ provided that $P_n \leq p_*$, where $p_*$ is a user defined small probability value. This simply states that, with probability $1-p_*$, the searcher is certain that the target is not within the radius $L_0$. This allows us to determine $n$. From
\begin{equation}
p_* = [1-\exp(-L_0/a)]^n
\end{equation}
we have:
\begin{equation}
    n = \ln(p_*)/\ln(1-\exp(-L_0/a)).
\end{equation}
The parameter $\tau_d$ is then $\tau_d=n\tau_0$, where $\tau_0$ is the sensing time introduced in Sec.\ref{s:prob}.

It remains to explain how the searcher to choose the position to jump to in the ballistic mode. First it is necessary to propose (generate) an action set $\mathcal{A}_k$. An action $\alpha\in \mathcal{A}_k$ consists  of a distance and angle pair $(L,\varphi)$, relative to the current searcher position. The cardinality of the action set $A = |\mathcal{A}_k|$, is a user defined parameter. A random sample $\{t\}_{1\leq j\leq A}$ is drawn according to (\ref{e:exp}) from the exponential distribution with parameter $\tau=\tau_b$. For each proposed $t_j$, the flight distance is computed as $L_{j} = t_j\,V_0$. The associated angle is a random draw from the uniform distribution, i.e.  $\varphi_j\sim \mathcal{U}[0,2\pi]$.
For each proposed action $\alpha\in \mathcal{A}_k$, a reward is computed via (\ref{e:Rew}). The action with the highest  reward is executed.

In diffusion mode, the searcher is static for the duration of   $\tau_d=n\tau_0$. After that, a ballistic flight is carried out, provided that the maximum probability in the threat map $\Pi_k = \max_m\{P_{k,m}; m=1,\dots,C\}$ is below a certain threshold $\zeta> 0.5$. Otherwise, the searcher will move to the nearest node in the grid  $\mathbf{G}$, closest to the cell whose probability of target presence equals $\Pi_k$. The speed of motion is $v_0\ll V_0$ (e.g.the searcher moves on water surface, without a takeoff). In this manner, the searcher will not fly away from the region if it suspects that the target may be present nearby.

Finally, the search is declared completed when the probability of target presence in one of the cells of the grid  $\mathbf{G}$ reaches a threshold $1-\epsilon$, where $\epsilon\ll 1$ is a user defined parameter.

\section{Numerical results}
\label{s:num}

\subsection{A single run demonstration}
Let us demonstrate the proposed search strategy with a  single run of the algorithm. The search area is described in Sec.~\ref{s:strategy}.A: $b=100$, $R_0=1$ a.u.. The target is located at coordinates $(35,60)$ a.u. Other parameters are: $a=3R_0$, $V_0=20$ a.u., $v_0=1$ a.u., $p_* = 0.05$, $\lambda=0.05$, $\zeta=0.7$, $\epsilon=10^{-3}$, $A=16$.

\begin{figure}[htb]
  \centerline{\includegraphics[width=8.5cm]{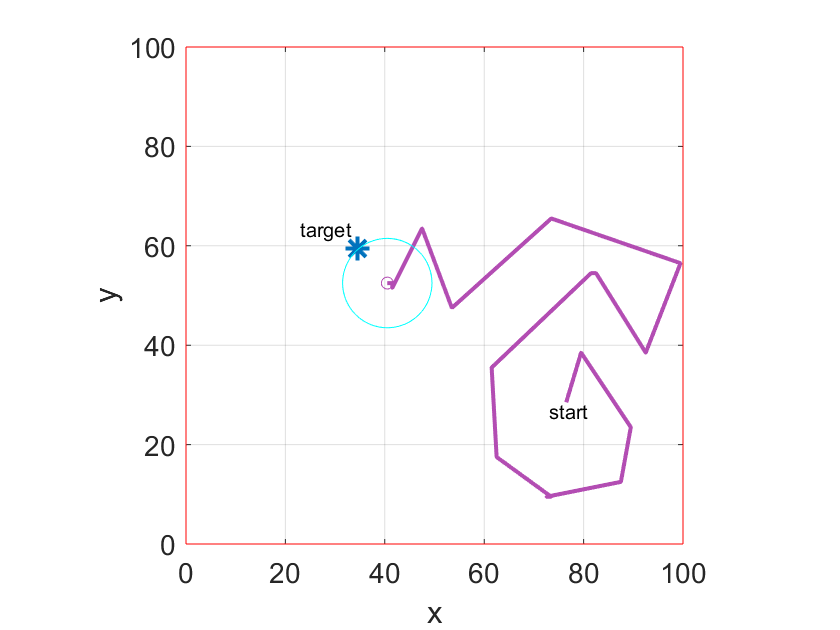} }
  \centerline{\footnotesize (a)}
  \centerline{\includegraphics[width=8.5cm]{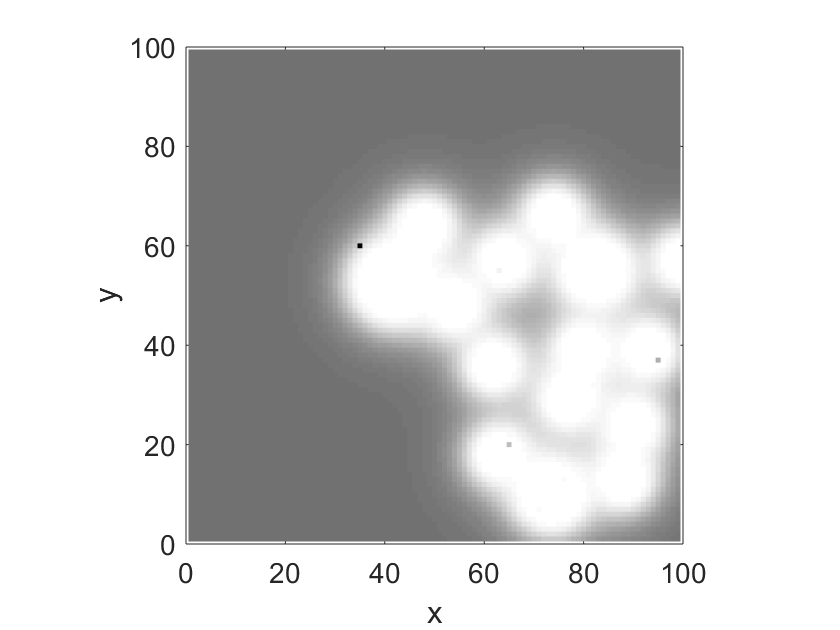} }
  \centerline{\footnotesize (b)}
  \centerline{\includegraphics[width=8cm]{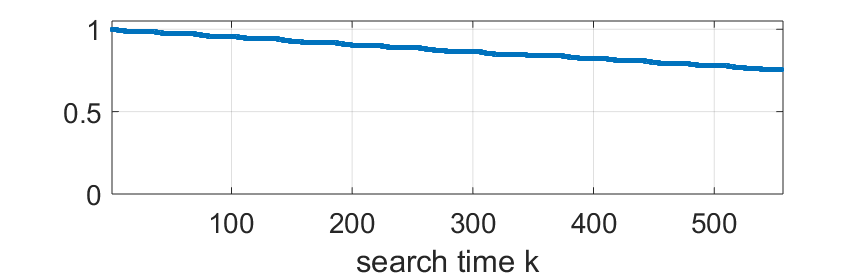} }
  \centerline{\footnotesize (c)}
  \caption{A demonstration of the proposed search strategy (a single run): (a) the search path; (b) the threat map at the end of the search period; (c) entropy $H_k$ versus time $k$.}
  \label{f:demo}
\end{figure}

Fig.~\ref{f:demo} shows (a) the search path, (b) the threat map at the end of the search period and (c) the evolution of entropy $H_k$ defined in (\ref{e:H}). The searcher started at a random location in the search area, and the target was found and correctly located after 555 a.u. of time. The cyan coloured circle in Fig.\ref{f:demo}.(a) indicates the sensing area $\mathcal{L}_k$.

\subsection{Monte Carlo runs}

\begin{figure}[htb]
  \centerline{\includegraphics[width=8cm]{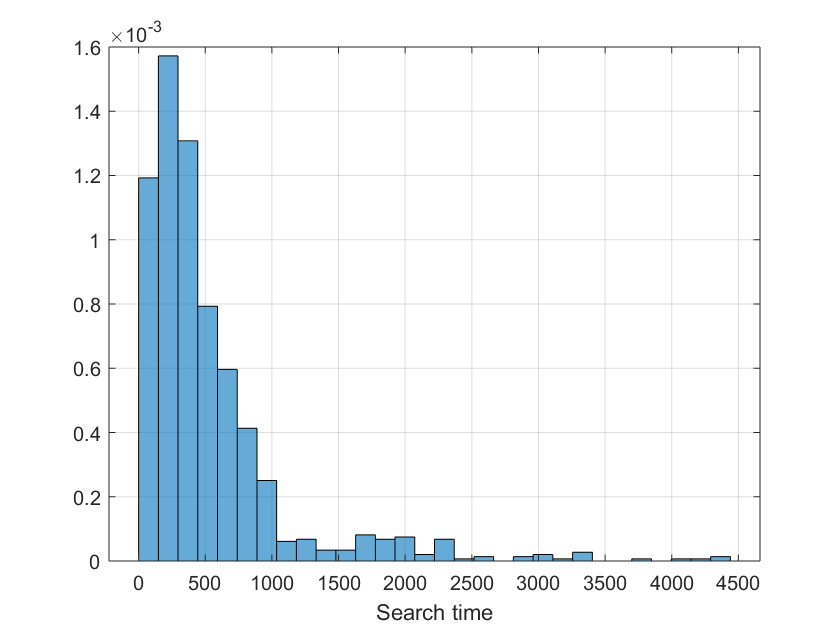} }
  \caption{A normalised histogram of search times ($a=3$)}
  \label{f:hist}
\end{figure}

The two most important performance criteria of any search algorithm are (i) the time required to find the source (the search time) and (ii) the success rate. We have carried out 1000 Monte Carlo runs of the proposed algorithm using the same parameters as listed above. The search failed to find the source within the period of 5000 a.u. only on 3 runs, resulting in success rate of $99.7$\%.  Fig.~\ref{f:hist}  shows the normalised histogram of the search time samples of 997 successful runs. The sample mean search time is $526.4$ a.u.  Observe that the  search time is characterised by a heavy-tailed distribution. A close inspection of this distribution reveals that it consists of two components, both of them inverse Gaussian (as postulated in \cite{skvortsov2018predicting}). The two components correspond to distinct phases of the search. In the first phase, the searcher jumps only to unexplored (virgin) regions of the search space before it finds it. In the second phase, the searcher has to re-examine the already explored areas, because the source has not been found in the first phase.

\begin{figure}[htb]
  \centerline{\includegraphics[width=8cm]{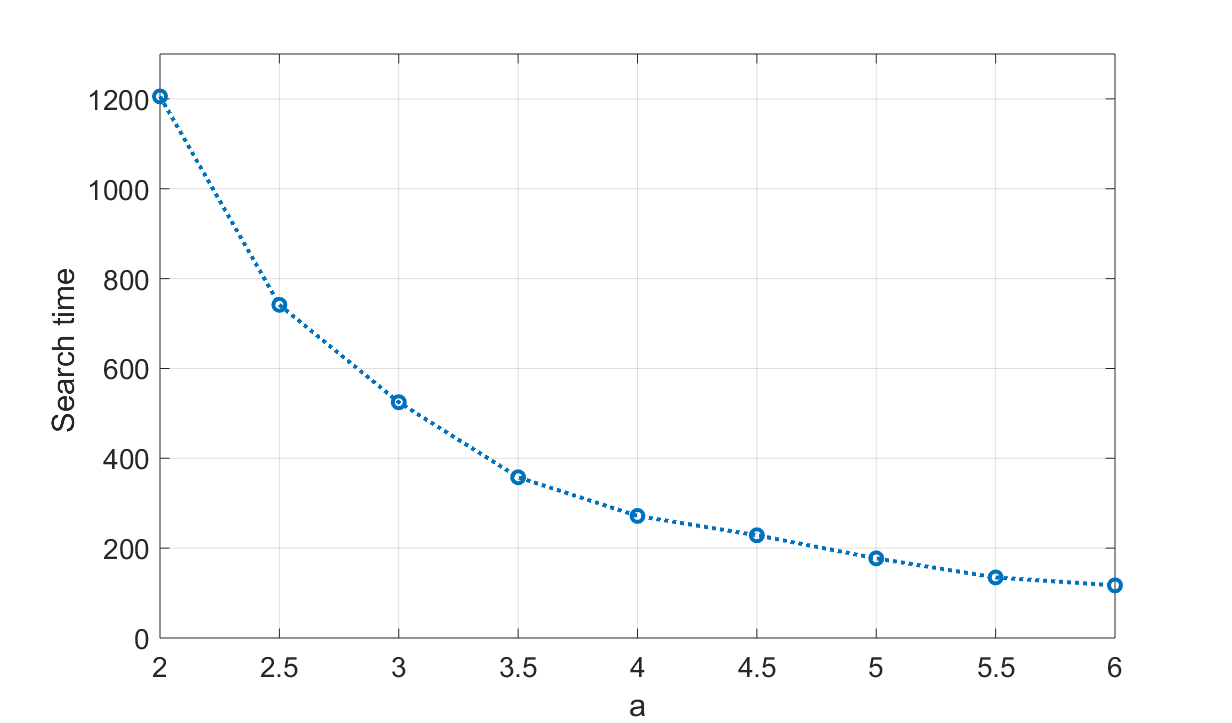} }
  \caption{Average search time as a function of sensing parameter $a$}
  \label{f:st}
\end{figure}

Next we repeat the described Monte Carlo runs, but using different values of the sensing parameter $a$. Fig.~\ref{f:st} shows the average search times as a function of parameter $a$. As expected, when the range of sensing (i.e. the values of $a$) is increased, keeping the search area $b^2$ constant, the time to find the source is shorter. The success rate for the considered range values of $a$ was always above $99.5$\%.

Finally, let us  contrast the proposed intermittent info-driven search with a pure info-driven search (infotaxi) strategy. The latter strategy, after some time, always creates a situation where equal reward (information gain) is assigned to all proposed actions at time $k$, $\alpha\in\mathcal{A}_k$. The searcher than becomes trapped in a local minimum of the threat map and the search never ends.

\section{Conclusions}
\label{s:conc}

The paper proposed a search strategy which combines the intermittent search with the information-driven search. The context is a search for targets using a sensor characterised by a probability of detection as a function of a distance to the target and a non-zero probability of false alarms. The parameters of the search strategy ($\tau_b$, $\tau_d$) are derived theoretically as a function of the sensor characteristic (parameter $a$), and the search volume (parameter $b$). Decisions, where to move, maximise the expected information gain.

There are two possibilities for future work. First would be to develop a scalable decentralised collaborative version of the proposed search strategy, for a swarm of interconnected UAVs. This approach would not require global knowledge of the communication network topology. Second direction of future work is theoretical: to determine the optimal value of parameter $p_*$ (which at present is user-specified) using the concept of information gain flux.

\end{document}